\documentclass[times,twocolumn,final,authoryear]{elsarticle}

\usepackage{jasr}
\usepackage{framed,multirow}

\usepackage{amssymb, amsmath}
\usepackage{tabularx}
\usepackage{rotating}
\usepackage{graphicx}
\usepackage{color, colortbl}
\definecolor{Gray}{gray}{0.9}

\usepackage{bm}

\usepackage{siunitx}

\usepackage{latexsym}

\usepackage{siunitx}

\def\araa{ARA\&A}            
\def\apj{ApJ}                
\def\apjl{ApJ}                
\def\apjs{ApJS}               
\def\aap{A\&A}                

\def\mnras{MNRAS}             
\def\ssr{Space~Sci.~Rev.}     

\usepackage{url}
\usepackage{xcolor}
\definecolor{newcolor}{rgb}{.8,.349,.1}

\def\delV{\Delta\mathrm{V}}

\def\infV{\mathrm{V}_\infty}

\def\changeformat#1{\xchangeformat#1\relax}
\def\xchangeformat#1/#2/#3\relax{%
#3 %
\ifcase#2 \or
JAN\or FEB\or MAR\or APR\or MAY\or JUN\or JUL\or
AUG\or SEP\or OCT\or NOV\or DEC\fi
\ #1}

\usepackage[citebordercolor=white]{hyperref}
\journal{Advances in Space Research}

\begin{document}

\verso{Andreas M. Hein \textit{et al}}

\begin{frontmatter}

\title{Interstellar Now!
Missions to Explore Nearby Interstellar Objects}%

\author[1]{Andreas M. Hein\corref{cor1}}
\ead{andreas.hein@i4is.org}
\cortext[cor1]{Corresponding author: Andreas M. Hein
  Tel.: +33680350667}
\author[2]{T. Marshall Eubanks}
\ead{tme@space-initiatives.com}
\author[3]{Manasvi Lingam}
\ead{mlingam@fit.edu}
\author[1]{Adam Hibberd}
\ead{adam.hibberd@ntlworld.com}
\author[1]{Dan Fries}
\ead{dan12fries@web.de}
\author[4]{Jean Schneider}
\ead{jean.schneider@obspm.fr}
\author[6]{Pierre Kervella}
\author[1]{Robert Kennedy}
\author[5,1]{Nikolaos Perakis}
\ead{nikolaos.perakis@tum.de}
\author[7]{Bernd Dachwald}

\address[1]{Initiative for Interstellar Studies (i4is) 27/29 South Lambeth Road London, SW8 1SZ United Kingdom}
\address[2]{Space Initiatives Inc, Palm Bay, FL 32907, USA}
\address[3]{Department of Aerospace, Physics and Space Sciences, Florida Institute of Technology, Melbourne FL 32901, USA}
\address[4]{Observatoire de Paris - LUTH, 92190 Meudon, France}
\address[5]{Department of Space Propulsion, Technical University of Munich, Germany}
\address[6]{LESIA, Observatoire de Paris, Université PSL, CNRS, Sorbonne Université, Université de Paris, 5 place Jules Janssen, 92195 Meudon, France}
\address[7]{Faculty of Aerospace Engineering, FH Aachen University of Applied Sciences, 52064 Aachen, Germany}

\received{1 May 2013}
\finalform{10 May 2013}
\accepted{13 May 2013}
\availableonline{15 May 2013}
\communicated{S. Sarkar}

\begin{abstract}
The recently discovered first high velocity hyperbolic objects passing through the Solar System, 1I/'Oumuamua and 2I/Borisov, have raised the question about near term missions to Interstellar Objects. \textit{In situ} spacecraft exploration of these objects will allow the direct determination of both their structure and their chemical and isotopic composition, enabling an entirely new way of studying small bodies from outside our solar system. In this paper, we map various Interstellar Object classes to mission types, demonstrating that missions to a range of Interstellar Object classes are feasible, using existing or near-term technology. We describe flyby, rendezvous and sample return missions to interstellar objects, showing various ways to explore these bodies characterizing their surface, dynamics, structure and composition. Interstellar objects likely formed very far from the solar system in both time and space; their direct exploration will constrain their formation and history, situating them within the dynamical and chemical evolution of the Galaxy. These mission types also provide the opportunity to explore solar system bodies and perform measurements in the far outer solar system.\\
\end{abstract}


\end{frontmatter}


\section{Introduction} \label{SecIntro}
It is not an exaggeration to contend that we live in a special epoch in which, after centuries of speculation, the first exoplanets have been detected \citep{Perr18}. One of the most compelling reasons for studying exoplanets is that discerning and characterizing these worlds holds the potential of revolutionizing our understanding of astrophysics and planetary science, as well as astrobiology if they are determined to harbor ``alien'' life \citep{SKP18,LL19}. Terrestrial telescopes and even futuristic very large instruments in space, such as the Labeyrie Hypertelescope \citep{Labeyrie-2016-a}, will not be sufficient to characterize and understand local geology, chemistry and possibly biology of extrasolar objects at small scales. Even minuscule gram-scale probes, laser-launched from Earth to relativistic speeds, are unlikely to return science data from other stellar systems much sooner than 2070 \citep{PSG16,Hein-et-al-2017-b,Park18,HKC19}.

However, these are not the only two avenues open to humanity. Extrasolar objects have passed through our home system twice now in just the last three years: 1I/'Oumuamua and 2I/ Borisov \citep{Meech-et-al-2017-a,Jewitt-and-Luu-2019-a}. These interstellar objects (ISOs) provide a previously inaccessible opportunity to directly, and much sooner, sample physical material from other stellar systems. By analyzing these interstellar interlopers, we can acquire substantial data and deduce information about their planetary system of origin \citep{FJ18,PTP18,MM18,JTH18}, planetary formation \citep{TRR17,RAV18,RL19}, galactic evolution, and possibly molecular biosignatures \citep{Lingam-Loeb-2018-a} or even clues about panspermia \citep{Ginsburg-et-al-2018-a}.

Previous papers have investigated flyby missions to ISOs either in the context of specific objects such as 1I/'Oumuamua and 2I/Borisov \citep{Hibberd2019,Hein-et-al-2017-a,HHE20} or objects passing through the Solar system which have been discovered early enough \citep{Seligman-Laughlin-2018-a,MCF20}. Furthermore, the Comet Interceptor mission, which has been selected as a Fast Class mission by the European Space Agency (ESA), aims to intercept a long-period comet \citep{SJ19}. In case this primary objective is not fulfilled, alternative candidates include not just short-period comets but also interstellar objects (ISOs) \citep{SKJS}, provided that the appropriate Delta-V requirements are met for the latter. Furthermore, sample return missions to outer solar system objects have also been proposed, which may face similar challenges as chasing ISOs \citep{mori2020solar}. In contrast, during the course of this paper, we map various ISO classes to mission types, demonstrating that missions to a range of ISO classes are feasible, all through the usage of existing or near-term technology. 

\section{ISO Mission Science Objectives}

ISOs passing through the solar system are the only interstellar objects we have a chance of directly exploring in the near future. The Rosetta mission has illustrated the limits of remote astronomical observations in characterizing a cometary body, and what can be achieved through direct exploration \citep{Drozdovskaya-et-al-2019-a}. Extending \textit{in situ} spacecraft exploration to ISOs ought to enable the determination of their surface, structure, and chemical and isotopic composition in detail. Initial studies have shown the existence of possible missions, solely reliant on existing technology, to 1I and 2I \citep{Hein-et-al-2017-a,Hibberd2019,HHE20}, and to additional, yet-to-be-discovered, ISOs \citep{Seligman-Laughlin-2018-a,MCF20}. In addition, work is ongoing on interstellar precursor missions deep into the outer solar system \citep{Brandt-et-al-2017-a,HAH20}, which are distinguished by similar trajectories \citep{McAdams-McNutt-2020-a} and could therefore be sent to intercept ISOs as a secondary objective.

A mission designed to target ISOs can yield valuable scientific return both prior to and after interception. Before the encounter, the probe could analyze interplanetary dust \citep{GGD01} and the solar wind plasma \citep{BC13}. Furthermore, much like the \emph{Spitzer} telescope, the mission may be suitable for microlensing studies, which yield information about the mass, distance and parallax vector of extrasolar objects \citep{UYG15,ZCG16}. 

\subsection{ISO Taxonomy}

To date, two different ISOs have been discovered. Their observed properties vary substantially: the hyperbolic interstellar asteroid (1I/'Oumuamua) and interstellar comet (2I/Borisov). Hyperbolic visitors that will not return to the Solar system are readily classifiable in terms of their composition and excess velocity at infinity (v$_\infty$); furthermore these parameters may exhibit some degree of correlation \citep{Eubanks-2019-b,Eubanks-2019-c}. We can reasonably expect other interstellar objects in the coming years, especially as astronomical surveys improve \citep{TRR17,PTP18,RL19,Yeh-et-al-2020-a}. In addition, captured ISOs in our solar system might already exist \citep{Torbett-1986-a,Gaidos18,Lingam-Loeb-2018-a,Siraj-Loeb-2018-a,Namouni-Morais-2018-a} -- some with very low original excess velocities that readily facilitated capture \citep{Belbruno-et-al-2012-a,Hands-Dehnen-2020-a,PDK20} -- although \citet{Morbidelli-et-al-2020} have challenged this origin for Centaurs. 

Distinguishing ISOs passing through our Solar system can be done via dynamical considerations, e.g., by measuring their speeds and thereby calculating their excess velocities; this method is valuable for objects with $v_\infty$ of at least a few km/s. In the case of captured ISOs, there are more ambiguities surrounding their existence and means of detecting them. If captured ISOs do exist in significant numbers, they may be discernible through their orbital parameters (especially the inclination), although the estimates vary from study to study \citep[cf.][]{Siraj-Loeb-2018-a,Namouni-Morais-2018-a,Hands-Dehnen-2020-a,Morbidelli-et-al-2020}. Furthermore, if the isotopic ratios (e.g., of the three oxygen isotopes) of putative captured ISOs diverge significantly from those of Solar system values, that would provide another means of distinguishing them \citep{Gaidos18,Lingam-Loeb-2018-a}. Further measurements of isotopic ratios and other chemical properties of ISOs passing through our Solar system will enable us to gain a better understanding of their properties, which may then be utilized in identifying captured ISOs. On account of the inherent uncertainties concerning captured ISOs, we emphasize that the entries (6) to (9) delineated in Table \ref{tab:ISOtypes} as well as the specific examples investigated in more detail such as Ka'epaoka'awela (514107) should be regarded as \emph{potential} candidates; in other words, these objects have not been unequivocally confirmed to be captured ISOs.

In order to truly determine whether objects are truly ISOs or not, visiting them is of paramount importance. Type 2 ISOs in Table \ref{tab:ISOtypes}, with v$_\infty$ $\lesssim$ 1 km/s, have been separately classified because the problem lies not just in finding but also in distinguishing them from long-period Oort Cloud comets (see, e.g., \citealt{Belbruno-et-al-2012-a,Krolikowska-et-al-2013-a,Hands-Dehnen-2020-a}). Ultimately, settling this issue necessitates compositional and isotopic analysis, which can be performed in fast flybys sampling the coma directly or collecting ejecta from impactors (see \citealt{Eubanks-et-al-2020-b}). The different mission categories and accompanying objectives are described in more detail in following sections.

\begin{table*} 
\centering
\caption{The InterStellar Object Taxonomy; types of ISOs, the associated science and potential near-term mission types. All missions, and especially rendezvous or sample return missions, are facilitated for ISOs having low inclinations, low v$_\infty$, and  for ISOs discovered well before their perihelion passage.}
\label{tab:ISOtypes}
\vspace{0.1 in}
\begin{tabular}{|l|p{6cm}|p{4.2cm}|p{5.5cm}|}
\hline 
ID & Type & Examples & Mission Type\tabularnewline
\hline 
\hline 
(1) & Clearly hyperbolic galactic thin disk objects, which have  1 km/s $\ll$ v$_\infty$ $\lesssim$ 100 km/s relative to the Sun.  & 1I/'Oumuamua \& 2I/Borisov. Currently detection limited. $\sim$92\% of arrivals in kinematic model.
& Flyby/Impactor, in fortuitous cases (especially with early pre-perihelion detection) rendezvous or sample return.   \\
\hline 
(2) & Galactic thick disk objects, with lower spatial density and higher velocities, roughly 100 $\lesssim$ v$_\infty$ $\lesssim$ 200 km/s. & None known so far. $\sim$6\% of arrivals in kinematic model. & Only flyby missions are possible, and only if discovered before perihelion. \\
\hline 
(3) & Galactic halo objects, with an even lower spatial density and v$_\infty$ $\gtrsim$ 200 km/s. & None known so far. $\sim$1\% of arrivals in kinematic model. & Probably not feasible even for flybys; would pass through the Earth's orbit in a few weeks or less. \\
\hline 
(4) & Bodies not bound to our galaxy. An very low spatial density and a galactic velocity  $\gtrsim$ 530 km/s. & None known so far. $\sim$0.4\% of arrivals in kinematic model. & Probably not feasible even for flybys; would pass through the Earth's orbit in less than 1 week. \\
\hline 
(5) & Similar to Type 1 objects, but with $v_\infty \lesssim 1$ km/s. This category is separated as these objects may be confused with the ``Oort spike'' of long period comets. & C/2007 W1 Boattini. Note that apparent interstellar comets at these velocities may be recaptured  Oort cloud comets.  & Flyby/Impactor/Rendezvous/Sample return \\
\hline 
(6) & Comets captured in the Oort cloud at the formation of solar system, and later perturbed into the inner system with other long period comets. & Population unknown, possibly a significant fraction of the long period comets.& Impact sampling or sample return, isotope analysis needed for confirmation. \\
\hline 
(7) & Objects (planetesimals) captured primordially by gas drag in early inner solar system. & Unclear if any has survived until now. & Rendezvous depending on inclination. Distinguishing them remotely will be hard.\\
\hline 
(8) & Captured objects in retrograde and other unusual orbits; see, e.g., \citet{Siraj-Loeb-2018-a,Namouni-Morais-2018-a,Morbidelli-et-al-2020}. These orbits are typically not stable and so these objects would be relatively recent captures. & Some Centaurs; retrograde objects such as (514107)~Ka'epaoka'awela. Work is needed to find orbits most likely to contain ISOs. & These objects are now in solar orbits, and rendezvous or sample return is possible  depending on their inclination.\\
\hline 
(9) & Sednoids, three body traded objects, special case of case \#6 or case \#8. The difference is that these objects are thought to have been captured in a 3-body interaction with the Sun and a passing star or planet.  & Sedna, 2014~UZ224, 2012~VP113, 2014~SR349, 2013~FT28 & Large distances, but low velocities, would facilitate rendezvous or sample return.\\
\hline 
\end{tabular}
\medskip
\end{table*}

\subsection{Overview of mission and science objectives}

There are three broad mission categories that naturally come to the fore, and their scientific potential (along with accompanying pros and cons) is described in more detail below. Unless explicitly noted otherwise, it may be assumed that all scientific objectives possible for a simpler mission can also be accomplished by more complex missions.

An \textbf{ISO flyby} provides opportunities for close-up observations and surface characterization as well as sample collection, either from the object’s plume or coma (for an active comet), or by liberating material through an impactor(s). Assuming a hypervelocity impact, radiation and detritus from the ionized plume could be analyzed using a high resolution UV spectrometer or mass spectrometer \citep{1991RCMS....5..441M,TMAR,Eubanks-et-al-2020-b}. Recommended strike velocities are in the narrow range of $3$-$6$ km/s; higher velocities could lead to over-fragmentation of biomolecular building blocks, whereas lower velocities render the method ineffective \citep{Klenner-et-al-2020}. Collected samples can be analyzed in flight by means of an onboard mass spectrometer, yielding information about composition and isotope ratios \citep{NKP20}. For more massive ISOs, detailed spectroscopic measurements of the target could yield further clues about the object’s composition and potentially even its history and origin in the galaxy. For instance, if the ratio of $^{12}$CO/$^{13}$CO is higher than the local interstellar medium value, it may indicate that the ISO in question spent a significant fraction of time in the vicinity of solar-type Young Stellar Objects \citep{SPY15}. Oxygen isotope ratios are also heterogeneous in different regions of the Galaxy \citep[e.g.][]{NG12}, and might therefore be indicative of where the ISO had originated.

An \textbf{ISO rendezvous with an Orbiter} would provide scientists with significantly more time for an in-depth and close-up study with a suite of instruments on board the orbiter or lander; analogous to, e.g., the Dawn and Rosetta missions \citep[e.g.,][]{GBKK07,RMJ15,TAB17}. Besides the object’s mass, density, mass distribution and composition, such a mission could perform seismologic experiments unveiling the deep interior structure of the ISO. Mass, density and crystalline structure (via microscopy) may be potentially determined for near-surface materials. Detailed measurements made possible by this type of mission might also yield information regarding the evolution of the originating stellar system. Depending on the instrumentation onboard the spacecraft, spectrophotometric, magnetometric, and radio measurements can be executed. Additionally, an \textbf{ISO rendezvous including a lander} could exploit advances in miniaturizing diagnostic equipment (e.g. lab-on-a-chip) and leverage the capabilities of a lander to return a large amount of data about the ISO over an extended period of time to scientists on Earth, including but not limited to composition, and possible volatile and organic molecules; these putative landers could leverage existing concepts developed for the likes of Enceladus and Europa \citep{KFD15}. Since the interstellar object will subsequently leave the Solar System and perhaps pass through another planetary system, a lander as a technological object would be a signpost of our technological achievements for an alien ``civilization'', should one exist. It would represent an interstellar version of the '`Message from Earth'' on board Pioneer 10 \citep{sagan1972message}.

\textbf{ISO sample return} via high-velocity impacts is the most complicated and audacious strategy, akin to what was accomplished by the Genesis and Stardust missions \citep{BBB03,Bro14}. In general, this mission type would utilize available $\delV$ not to rendezvous, but to return back to Earth. Besides some of the aforementioned science objectives, returning samples to earth allows for much more detailed analysis essentially unconstrained by mass, size, resolving power, operating power, and time \citep{NAD20}. Molecular composition and micro-crystalline structure can be deduced from vaporised ejecta and dust. Determining mineralogic, mechanical and structural properties would need centimeter-sized samples, either collected in the plume/coma of the ISO or from ejecta generated by an impactor. Laboratories back on Earth could readily undertake analysis of the isotope ratios of heavy elements, molecular chemistry, nuclear chemistry, and neutron activity. Diagnostic equipment is self-evidently not subject to mass constraints of the spacecraft and can provide, among others, higher-resolution spectroscopy, spectrophotometry, electron- and atomic force microscopy. One trade-off is that a sample return mission may yield less information about basic mass, density and seismology of the target. Furthermore, with existing sample return technologies, the returned samples are limited to solid dust grains, which limits the understanding of comet-like objects containing volatiles.

\textbf{Additional non-ISO science objectives:} In addition to scientific objectives associated with the ISOs themselves, interesting measurements can also be performed en-route, including but not limited to the collection and analysis of interplanetary dust and ions and close-up observation of outer Solar system phenomena, e.g., the IBEX ribbon \citep{MZB17}. In case the mission is tailored toward a flyby of the ISO, it will continue on its prescribed trajectory and will eventually traverse and move beyond the heliosphere. In this process, it could yield a wealth of information about the heliosphere and interstellar medium (ISM), just as the \emph{Voyager} spacecraft do. 

Further scientific objectives include the shape of the heliosphere \citep{DKM17}, the propagation of galactic cosmic rays \citep{SCM13,CSH16}, and the interaction with the ISM \citep{Zank}. Some examples of ISM physics and characteristics worthy of further study are the radial large-scale gradient \citep{Kurth-Gurnett-2020-a}, interstellar plasma and magnetic fields \citep{GKB13,BN14}, and magnetic turbulence \citep{BFN15}. One concrete example of each of the three mission categories, outlined above, follows. We will not comment on the instrumentation, because it is not the thrust of this paper. Minimum instrumentation should, however, include a camera and mass spectrometer for each of the missions.

\section{Types of ISO Missions}\label{SecISO}

ISO missions can be characterized by the resources required to perform them, which are closely related to how the ISO came to be in the solar system, and whether a mission is able to interact with it before its perihelion or afterwards (see Table \ref{tab:Missions}). ISOs are either unbound, passing through the solar system on a hyperbolic orbit, or bound, in some elliptical orbit about the Sun or even one of the planets. Unbound ISOs will generally be clearly of interstellar origin, but will only pass through the solar system once. If a mission can be launched before or around the time of the ISO's perihelion passage, then travel times can be reduced, especially if the ISO passes close the Earth, and a fast sample return (capture of cometary coma or impact probe ejecta material) may be possible.

\begin{table*} 
\caption{Types of missions to InterStellar Objects}
\label{tab:Missions}
\vspace{0.1 in}
\begin{tabular}{|c|c|c|c|}
\hline 
Target & Mission Type & Exploration Type & Notes\tabularnewline
\hline 
\hline 
ISOs Entering the Solar System & Loiter Missions & Fast Sample Return
& Requires Prepositioning of Spacecraft \tabularnewline
\hline 
ISOs Leaving the Solar System & Chase Missions & Fast Flyby & High $\delV$, Long Duration \tabularnewline
\hline 
Captured ISOs & Preplanned Missions & Orbiters, Landers & Similar to other asteroid/comet missions.  
\tabularnewline
\hline
\end{tabular}
\medskip
\end{table*}

\subsection{Flyby Missions}\label{SecFly}

Flyby missions are of particular relevance where the distance of the target ISO from the sun is large and/or the ISO is travelling at a high heliocentric speed. The former may be for one of the following two reasons:
\begin{enumerate}
    \item The perihelion is high, so minimum possible encounter distances are still extremely large (e.g. type (7) ISOs in Table \ref{tab:ISOtypes}).
    \item The perihelion is small, but the detection of the ISO occurs too late to take full advantage of this fact; for example, the type (1) ISO, 1I/‘Oumuamua, in Table \ref{tab:ISOtypes}.
    \end{enumerate}

\begin{figure}
\centering
\includegraphics[scale=0.36]{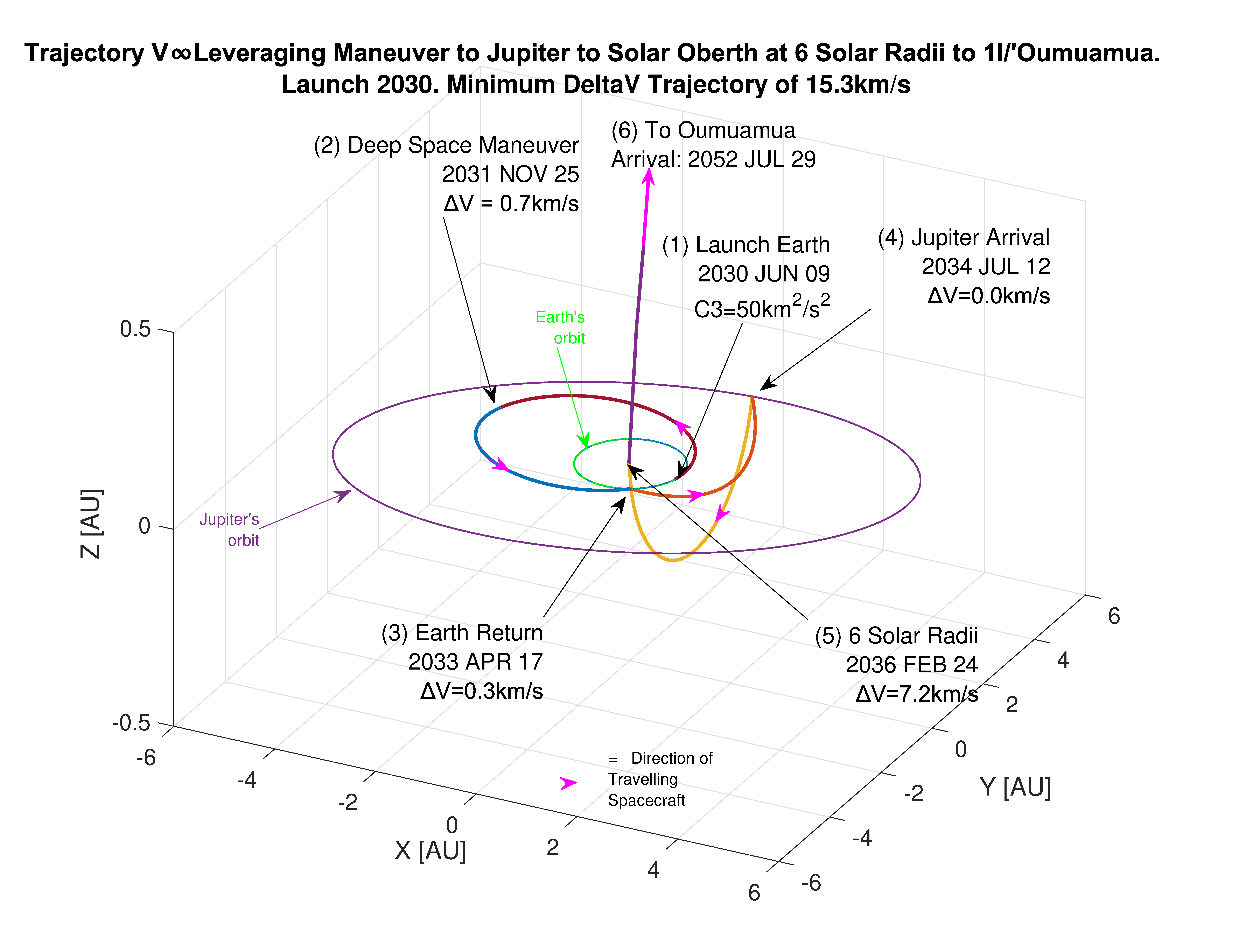}
\caption{Trajectory to 1I/'Oumuamua}
\label{fig:EDSMEJ6SR1I}
\end{figure}

A high intercept distance means that a large sun-radial velocity component must be generated in order to constrain the flight duration to a practically acceptable value. For chemical propulsion to 1I, extensive research has been conducted \citep{Hein-et-al-2017-a,HHE20}. The mission shown in Figure \ref{fig:EDSMEJ6SR1I} is a launch in 2030 and a `$\infV$ Leveraging Maneuver', a reverse gravity assist (GA) at Jupiter, followed by a Solar Oberth (SO) maneuver at 6 solar radii \citep{Blanco-Mungan-2020-a}, and 2-stage sample return mission at the SO which enables intercept at $200$\si{AU}. Using the Space Launch System (SLS), depending on the version, a probe mass up to $\sim 900$ \si{kg} is possible. More generally, launchers such as the Falcon Heavy and SLS can be used to throw spacecraft with masses up to $1000$ kg to ISO targets depending on launch date, mission duration, and maneuvers \citep{Hein-et-al-2017-a,HHE20}. 

For the SO maneuver, at 6 solar radii, heatshield technology similar to the Parker Solar Probe can be used to protect against solar heating \citep{HHE20,Brandt-et-al-2017-a}. Due to uncertainty in 1I’s orbit, at 200\si{AU} there is a possible displacement on the order of $10^{5}$ \si{km} from its estimated solar escape asymptote, assuming a positional uncertainty of $10^{-5}$ \si{rad} \citep{trilling2018spitzer}. At an approach speed of 30 \si{km.s^{-1}}, observations from the spacecraft would require a New Horizons LORI-type telescope (apparent magnitude of 17 at 10 \si{s} exposure time \citep{cheng2009long}. Assuming an apparent magnitude of 26 of the object and 11 hours of exposure time, the object could be detected at a distance of about $4.6 \times 10^{6}$ \si{km}, which translates to a timescale of $43$ hours before closest approach for the specified speed of 30 \si{km.s^{-1}}.

The horizontal maneuver would require a velocity increment on the order of hundreds of \si{ms^{-1}}. As a more advanced approach, a swarm of chipsats could be dispensed around 1I’s estimated escape asymptote and travel in the vanguard of the probe, returning data which would allow the main craft to adjust its velocity accordingly to ensure intercept. The main craft would then release an impactor and analyze the isotopic composition of 1I via spectroscopic methods. However, the consequently smaller telescope size renders detection more challenging, as might the data return to Earth. The potential to sequentially launch the chipsats at velocities of 300 \si{kms^{-1}} or higher, such as with the Starshot precursor architecture \citep{Park18} may merit further research.

Our brief analysis (and its attendant caveats) should not be regarded as exhaustive. Other issues that we have not delineated include the difficulties posed by long CCD exposure times (11 hours in our scenario) such as the cumulative impact of cosmic rays and the necessity of accounting for parallax motion of the object during this period. Obstacles with respect to measuring the position of the object, calculating offsets, and relaying it to the spacecraft may also arise. Hence, we acknowledge that there are significant (but not necessarily insurmountable) and outstanding challenges that are not tackled herein, as they fall outside the scope of this particular paper.

For nuclear thermal propulsion (NTP) to 1I, \citet{Hibberd-Hein-2020-a} have shown that a direct trajectory leaving low Earth orbit (LEO) in 2030, to fly by 1I, is achievable using a small nuclear rocket engine (derived from the government-sponsored Rover/NERVA programs) and an SLS Block 2. Utilizing a Oberth maneuver at Jupiter to reach 1I drastically reduces flight time. Launching in 2031, a "Pewee"-class NTP system (also researched in the Rover/NERVA programs) can deliver 2.5 \si{t} on target in a 14 year flight. The flight segment from LEO to Jupiter would take 5 months and, thus, needs a zero-boil-off cryocooler and zero-leakage liquid hydrogen (LH2) tanks. Other existing/near-term technologies could also be applied to drastically reduce this mission‘s duration, e.g., solar sails, electric sails, and multi-grid electric thrusters  \citep{Dachwald-2004-a,Loeb-et-al-2008-a,Loeb-et-al-2011-a,Hein-et-al-2017-a,Brandt-et-al-2017-a}.

\subsection{Rendezvous Missions}
For \textbf{rendezvous missions to hyperbolic ISOs}, The high radial velocity required to achieve an acceptable flight duration, as discussed in Sec. \ref{SecFly}, would then need to be removed in order to achieve a rendezvous mission, thereby imposing severe constraints on the on-board propulsion system (hence the New Horizons flyby of Pluto for example). However, it should be noted that, although the technology would require some further research and development, rendezvous missions could utilise electric or magnetic sail propulsion schemes to slow down and stay with the target. Solar sails based on the statite concept have also been proposed as viable alternatives for rendezvous missions \citep{LLM20}. The specifics for a rendezvous mission were described for 1I/'Oumuamua as a target in \citet{Hein-et-al-2017-a}.

In a similar vein, it is instructive to further delve into a couple of representative examples for other ISOs, notably \textbf{rendezvous missions to captured ISOs}. Type (6) ISOs (Table \ref{tab:ISOtypes}) in elliptical orbits (as opposed to hyperbolic orbits of types (1) and (2)) follow periodic optima, and so can spacecraft. This opens the possibility of rendezvous missions with reasonable $\delV$ of approximately 10 km/s. Rendezvous missions require a thrust from the spacecraft as the target ISO is approached to slow down and stay with the ISO in its path around the sun. The two objects studied in more detail here are both potentially type (6) ISOs, namely (514107) Ka'epaoka'awela (which is in retrograde motion and co-orbital with Jupiter), and the highly inclined centaur 2008 KV42. 

\begin{figure}
\centering
\includegraphics[scale=0.36]{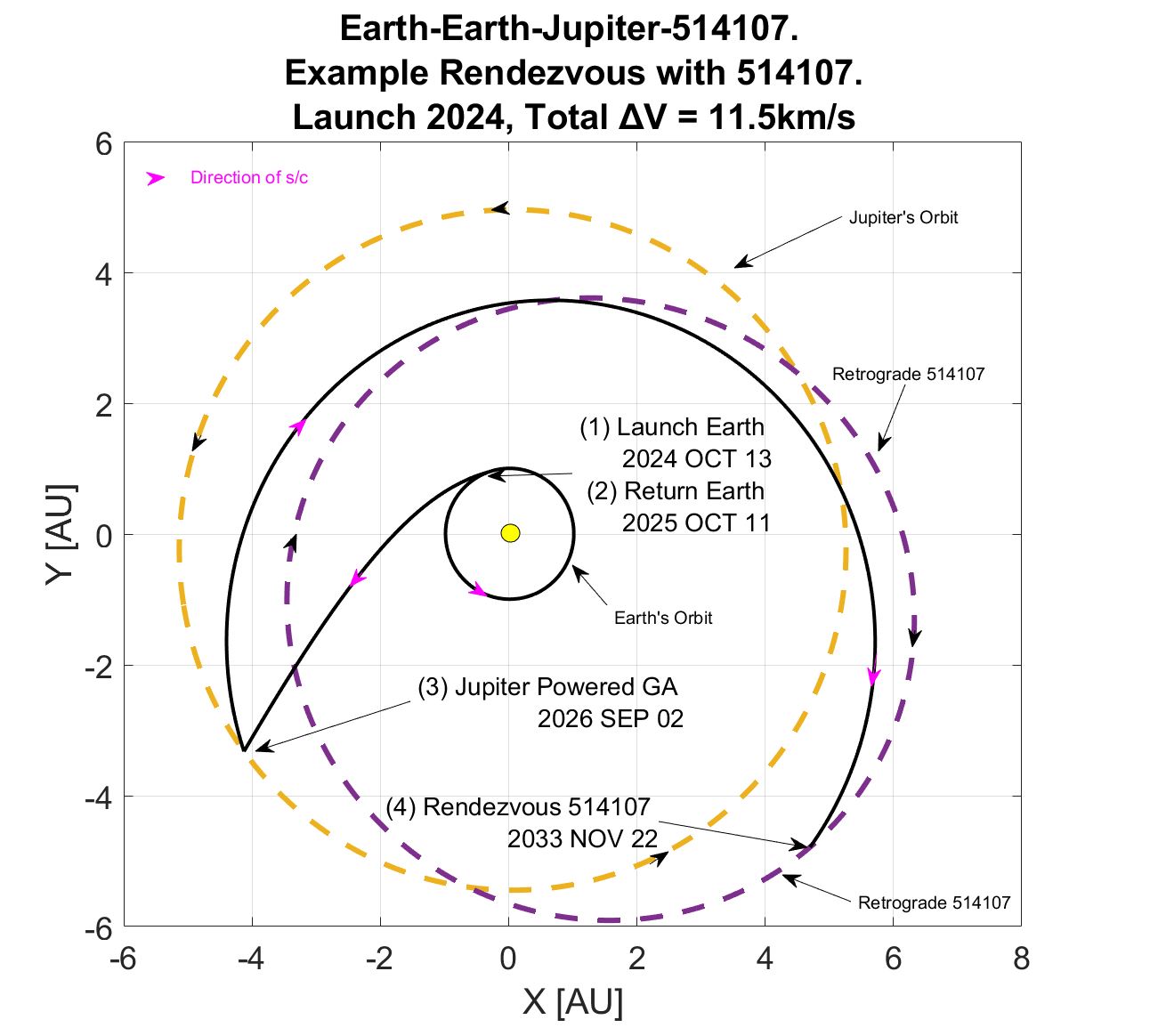}
\caption{Trajectory to Ka'epaoka'awela (514107) Launch 2024}
\label{fig:514107_1}
\end{figure}

\begin{figure*}
\centering
\includegraphics[scale=0.5]{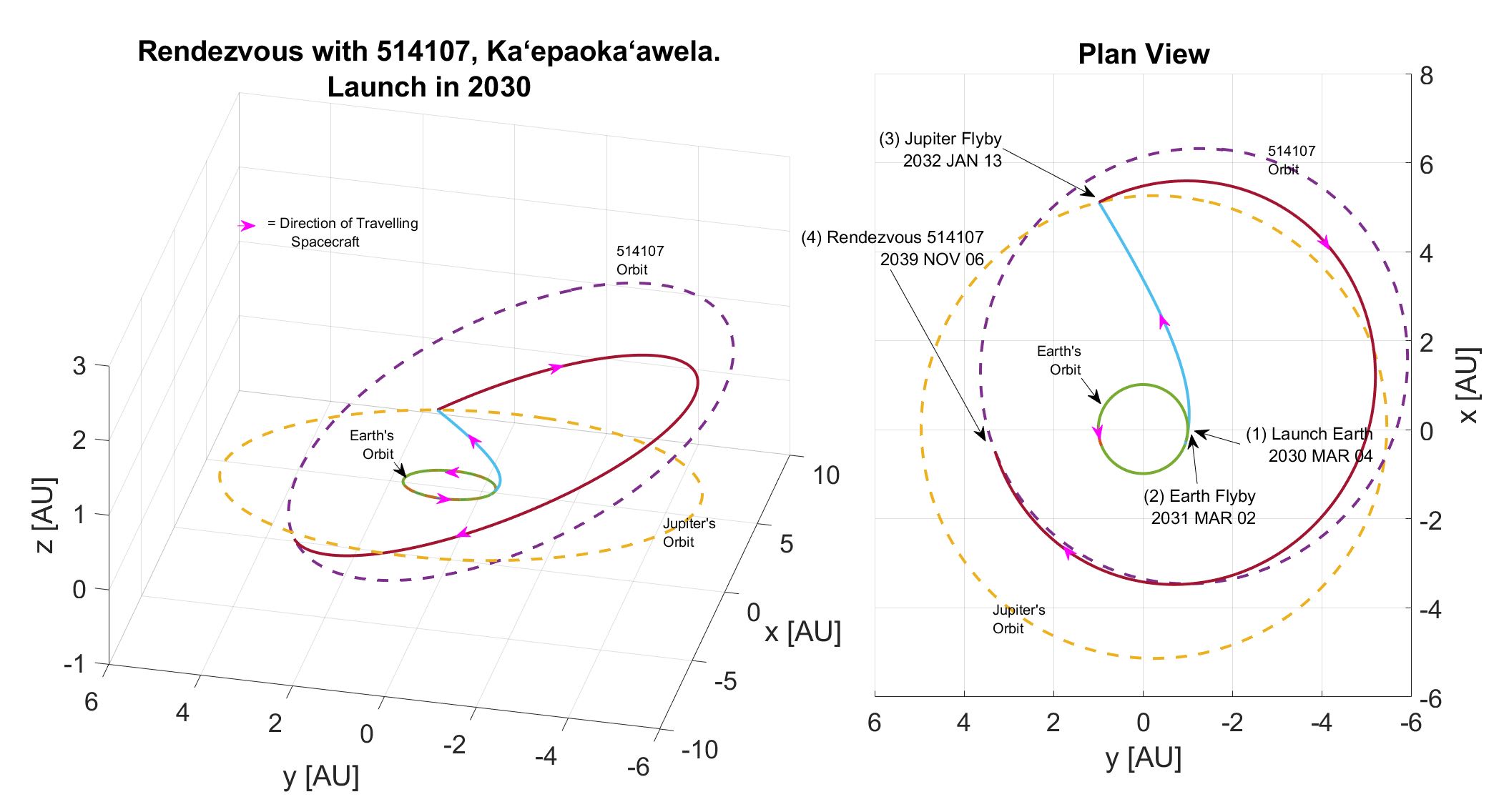}
\caption{Trajectory to Ka'epaoka'awela (514107) Launch 2030}
\label{fig:514107_2}
\end{figure*}

In the case of 514107, there are two relatively near-term rendezvous mission candidates launching in 2024 and in 2030. These are shown in Figures \ref{fig:514107_1} and \ref{fig:514107_2}, respectively. The latter opportunity has the advantage of a marginally lower $\delV$ and a later launch date to enable more time for mission preparation. The pertinent data is provided in Table \ref{table:514107Rendez}. Hence the launch is in 2030 with a $\infV$ Leveraging Manoeuvre of $n=1$ year. A Jupiter Oberth in Jan 2032 applied at an altitude of 77,198 \si{km} results in a retrograde heliocentric orbit. In this orbit the spacecraft travels on a long cruise, eventually catching up with 514107 and applying a $\delV$ of 2.5~\si{km.s^{-1}} to slow down and rendezvous. For completeness, the long spacecraft cruise arc from Jupiter to 514107 subtends an angle of \ang{272.6} at the sun.

\begin{figure}
\centering
\includegraphics[scale=0.33]{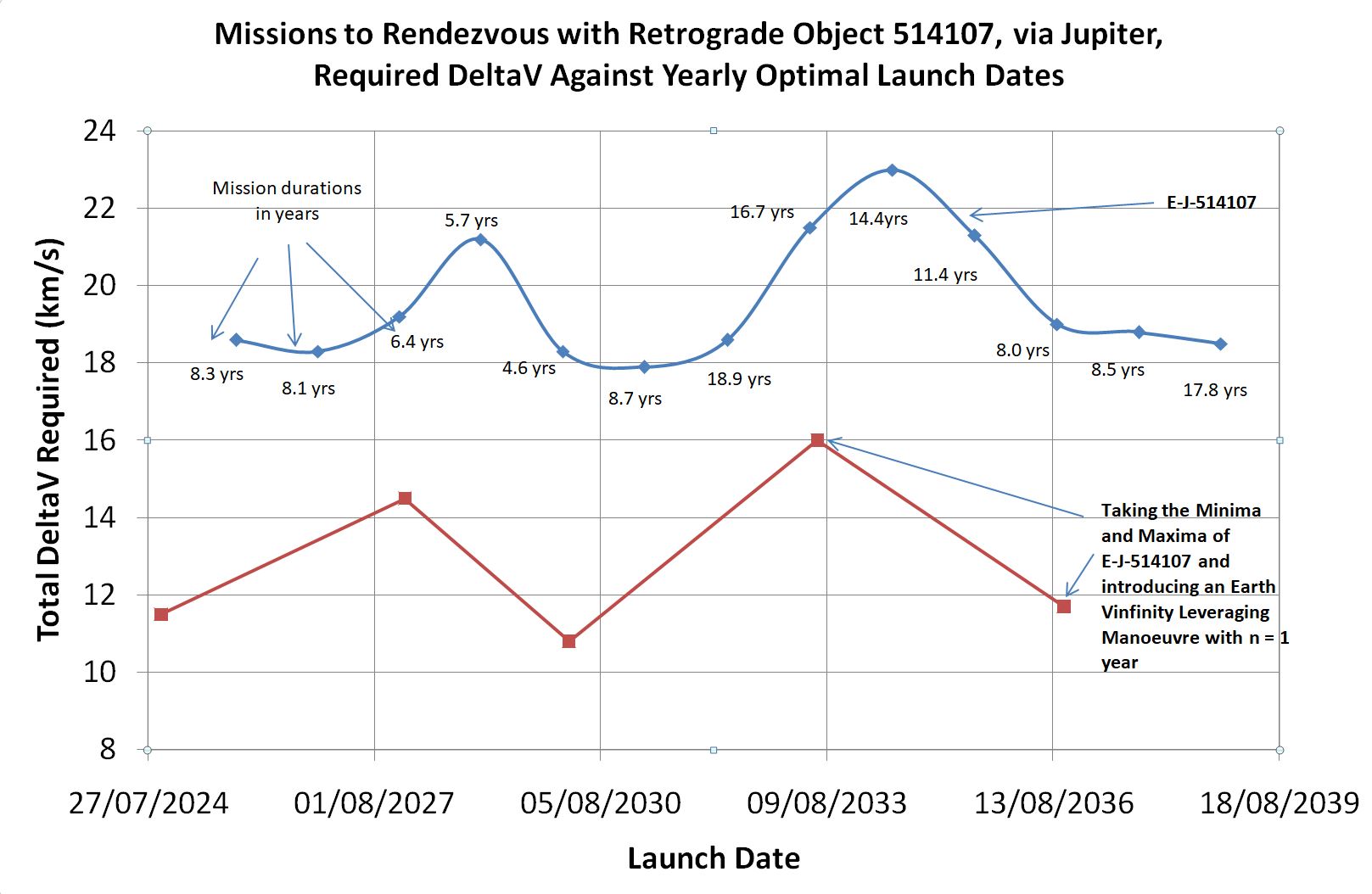}
\caption{$\delV$ Dependency on Launch Date for a Rendezvous Trajectory to Ka'epaoka'awela (514107)}
\label{fig:514107_3}
\end{figure}

\begin{figure}

\centering

\includegraphics[scale=0.27]{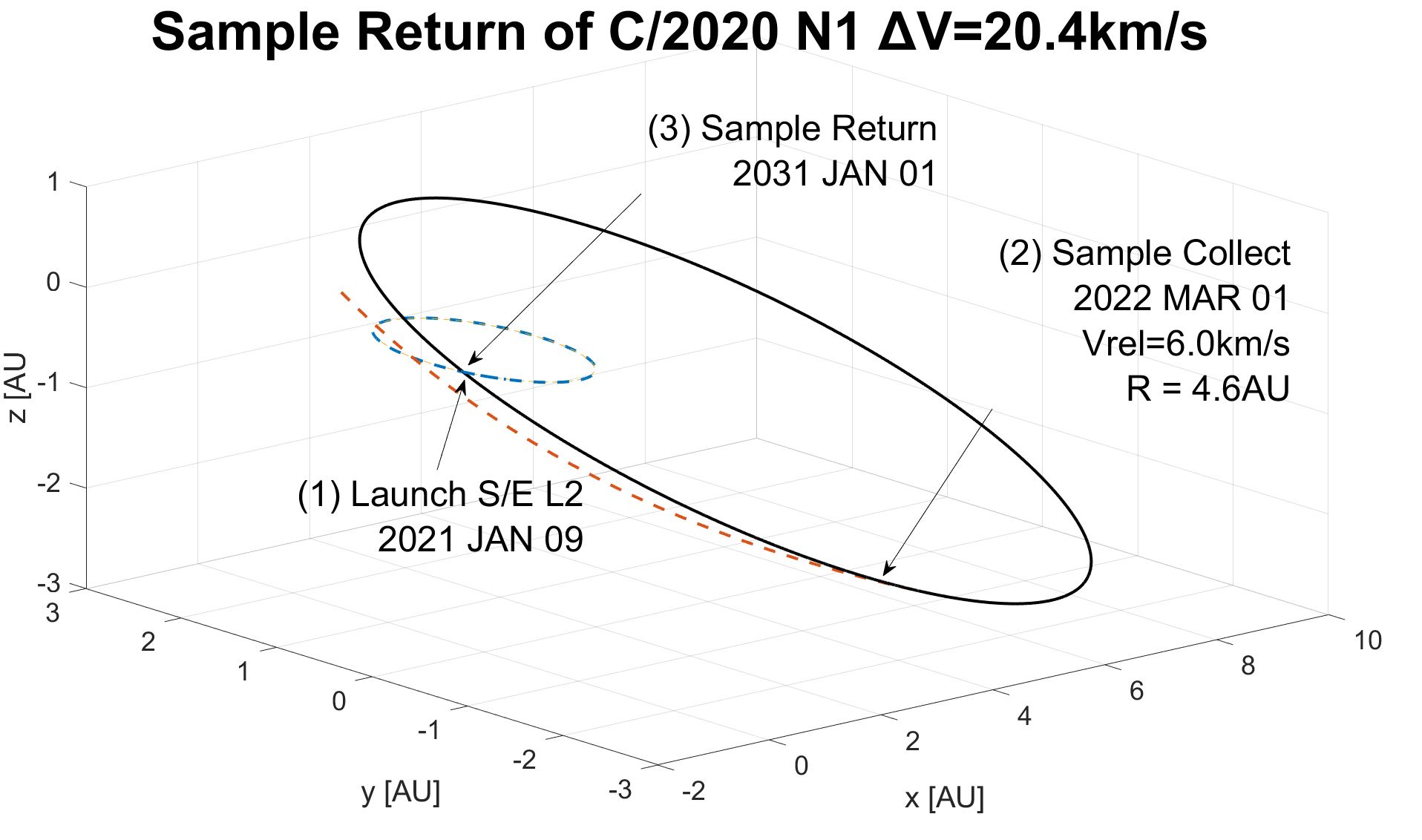}

\caption{Sample Return from C/2020 N1}

\label{fig:C2020N1}

\end{figure}

\begin{table*}[]

\caption{Sample Return from Several Weekly Hyperbolic Comets and Also 1I/'Oumuamua and 2I/Borisov}

\label{tab:samp_ret}

\resizebox{\textwidth}{!}{%

\centering

\begin{tabular}{|c|c|c|c|c|c|c|c|c|c|c|c|c|c|}

\hline

Object & Total & n & Discovery & Launch & Encounter & Return & R & Vrel & Vrel & Delta-V & Delta-V & Flight & Flight \\

 & Delta-V &  &  & from &  &  & encoun & encoun & return & at L2 & Object & D. & D. \\

 & km/s &  &  & S/E L2 &  &  & ter & ter & \si{km.s^{-1}} & \si{km.s^{-1}} & \si{km.s^{-1}} & \si{days} & \si{yrs} \\

 &  &  &  &  &  &  & \si{AU} & \si{km.s^{-1}} &  &  &  &  &  \\ \hline

C 2019 & 16.1 & n/a & \changeformat{28/12/2019} & \changeformat{04/03/2020} & \changeformat{31/05/2020} & \changeformat{10/04/2021} & 0.25 & 67.00 & 25.70 & 12.5 & 3.6 & 402 & 1.10 \\

Y4 Atlas &  &  &  &  &  &  &  &  &  &  &  &  &  \\ \hline

C 2020 & 6.7 & n/a & \changeformat{03/07/2020} & \changeformat{30/07/2020} & \changeformat{04/03/2021} & \changeformat{18/10/2021} & 1.32 & 19.80 & 11.60 & 3.40 & 3.30 & 445 & 1.22 \\

N1 P &  &  &  &  &  &  &  &  &  &  &  &  &  \\ \hline

C 2020 & 5.7 & n/a & \changeformat{03/07/2020} & \changeformat{03/07/2020} & \changeformat{04/03/2021} & \changeformat{20/10/2021} & 1.33 & 19.60 & 11.60 & 2.00 & 3.60 & 474 & 1.30 \\

N1 P &  &  &  &  &  &  &  &  &  &  &  &  &  \\ \hline

\rowcolor{Gray}

C 2020 & 20.4 & 10 & \changeformat{03/07/2020} & \changeformat{09/01/2021} & \changeformat{01/03/2022} & \changeformat{10/01/2031} & 4.55 & 5.9 & 20.6 & 20.4 & 0.0 & 3652 & 10.0 \\

\rowcolor{Gray}

N1 P* &  &  &  &  &  &  &  &  &  &  &  &  &  \\ \hline

C 2017 & 18.4 & 14 & \changeformat{30/09/2017} & \changeformat{18/02/2018} & \changeformat{31/10/2018} & \changeformat{20/02/2032} & 3.45 & 26.80 & 17.40 & 18.4 & 0.00 & 5115 & 14.00 \\

S6 &  &  &  &  &  &  &  &  &  &  &  &  &  \\ \hline

C 2018 & 8.5 & 6 & \changeformat{27/10/2018} & \changeformat{27/12/2019} & \changeformat{27/10/2021} & \changeformat{25/12/2025} & 5.00 & 23.00 & 8.90 & 8.5 & 0.00 & 2190 & 6.00 \\

U1 &  &  &  &  &  &  &  &  &  &  &  &  &  \\ \hline

C 2019 & 10.2 & 15 & \changeformat{29/03/2019} & \changeformat{07/05/2019} & \changeformat{18/01/2025} & \changeformat{07/05/2034} & 10.70 & 10.20 & 10.60 & 10.2 & 0.00 & 5479 & 15.00 \\

F1 Atlas &  &  &  &  &  &  &  &  &  &  &  &  &  \\ \hline

C 2014 & 9.1 & 3 & \changeformat{04/01/2014} & \changeformat{11/02/2014} & \changeformat{27/08/2015} & \changeformat{14/02/2017} & 2.90 & 20.70 & 7.00 & 6.00 & 3.10 & 1099 & 3.01 \\

AA 52 &  &  &  &  &  &  &  &  &  &  &  &  &  \\ \hline

\rowcolor{Gray}

C 2014 & 17.2 & 17 & \changeformat{16/12/2014} & \changeformat{30/03/2015} & \changeformat{27/01/2019} & \changeformat{04/04/2032} & 10.0 & 6.0 & 16.8 & 16.4 & 0.8 & 6215 & 17.0 \\

\rowcolor{Gray}

Y1* &  &  &  &  &  &  &  &  &  &  &  &  &  \\ \hline

C 2015 & 4.6 & 2 & \changeformat{11/01/2015} & \changeformat{30/01/2017} & \changeformat{04/07/2017} & \changeformat{31/01/2019} & 1.67 & 25.10 & 5.00 & 4.6 & 0.00 & 731 & 2.00 \\

V2 Johnson &  &  &  &  &  &  &  &  &  &  &  &  &  \\ \hline

C 2015 & 23.9 & 16.5 & \changeformat{20/05/2015} & \changeformat{28/05/2017} & \changeformat{29/11/2020} & \changeformat{15/12/2033} & 11.7 & 6.0 & 35.2 & 17.4 & 6.5 & 6045 & 16.5 \\

H2* &  &  &  &  &  &  &  &  &  &  &  &  &  \\ \hline

C 2013 & 16.3 & 12 & \changeformat{04/11/2013} & \changeformat{19/01/2014} & \changeformat{11/09/2015} & \changeformat{19/01/2026} & 5.80 & 9.20 & 16.40 & 16.3 & 0.00 & 4383 & 12.00 \\

V1 Boattini &  &  &  &  &  &  &  &  &  &  &  &  &  \\ \hline

C 2013 & 11.5 & 30 & \changeformat{04/11/2013} & \changeformat{05/01/2014} & \changeformat{19/09/2016} & \changeformat{06/01/2044} & 8.80 & 6.50 & 11.80 & 11.5 & 0.00 & 10958 & 30.00 \\

V1 Boattini &  &  &  &  &  &  &  &  &  &  &  &  &  \\ \hline

C 2018 & 7.5 & 3 & \changeformat{28/01/2018} & \changeformat{01/02/2018} & \changeformat{06/09/2018} & \changeformat{29/01/2021} & 2.30 & 15.20 & 7.30 & 7.5 & 0.00 & 1093 & 2.99 \\

C2 Lemmon &  &  &  &  &  &  &  &  &  &  &  &  &  \\ \hline

C 2018 & 18.7 & 6 & \changeformat{28/01/2018} & \changeformat{17/04/2018} & \changeformat{01/03/2019} & \changeformat{17/04/2024} & 3.67 & 10.0 & 18.6 & 18.6 & 0.1 & 2191 & 6.0 \\

C2 Lemmon &  &  &  &  &  &  &  &  &  &  &  &  &  \\ \hline

\rowcolor{Gray}

C 2018 & 24.4 & 14 & \changeformat{28/01/2018} & \changeformat{08/05/2018} & \changeformat{05/11/2019} & \changeformat{08/05/2032} & 5.8 & 6.0 & 24.2 & 24.2 & 0.2 & 5114 & 14.0 \\

\rowcolor{Gray}

C2 Lemmon * &  &  &  &  &  &  &  &  &  &  &  &  &  \\ \hline

C 2020 & 8.2 & 5 & \changeformat{25/05/2020} & \changeformat{30/06/2020} & \changeformat{14/06/2022} & \changeformat{30/06/2025} & 4.70 & 17.00 & 8.80 & 8.2 & 0.00 & 1826 & 5.00 \\

K5 PanSTARRS &  &  &  &  &  &  &  &  &  &  &  &  &  \\ \hline

2I Borisov & 6.2 & n/a & \changeformat{30/08/2020} & \changeformat{12/07/2018} & \changeformat{26/10/2019} & \changeformat{18/09/2020} & 2.20 & 33.00 & 12.20 & 5.00 & 1.20 & 799 & 2.19 \\ \hline

1I 'Oumuamua & 4.7 & 2 & \changeformat{19/10/2017} & \changeformat{23/07/2017} & \changeformat{24/10/2017} & \changeformat{19/07/2019} & 1.35 & 49.80 & 5.20 & 4.7 & 0.00 & 726 & 1.99 \\ \hline

\multicolumn{14}{|l|}{n indicates the number of years from launch to return of the sample.}\\

\multicolumn{14}{|l|}{Rows with * are missions with Encounter Vrel $<$ 6\si{km.s^{-1}}}\\

\multicolumn{14}{|l|}{ In addition gray missions have  Vrel $<$ 6\si{km.s^{-1}} and no $\delV$ at the object, and so n is an integer number of years}\\

\end{tabular}}

\end{table*}

\begin{figure*}
\centering
\includegraphics[scale=0.46]{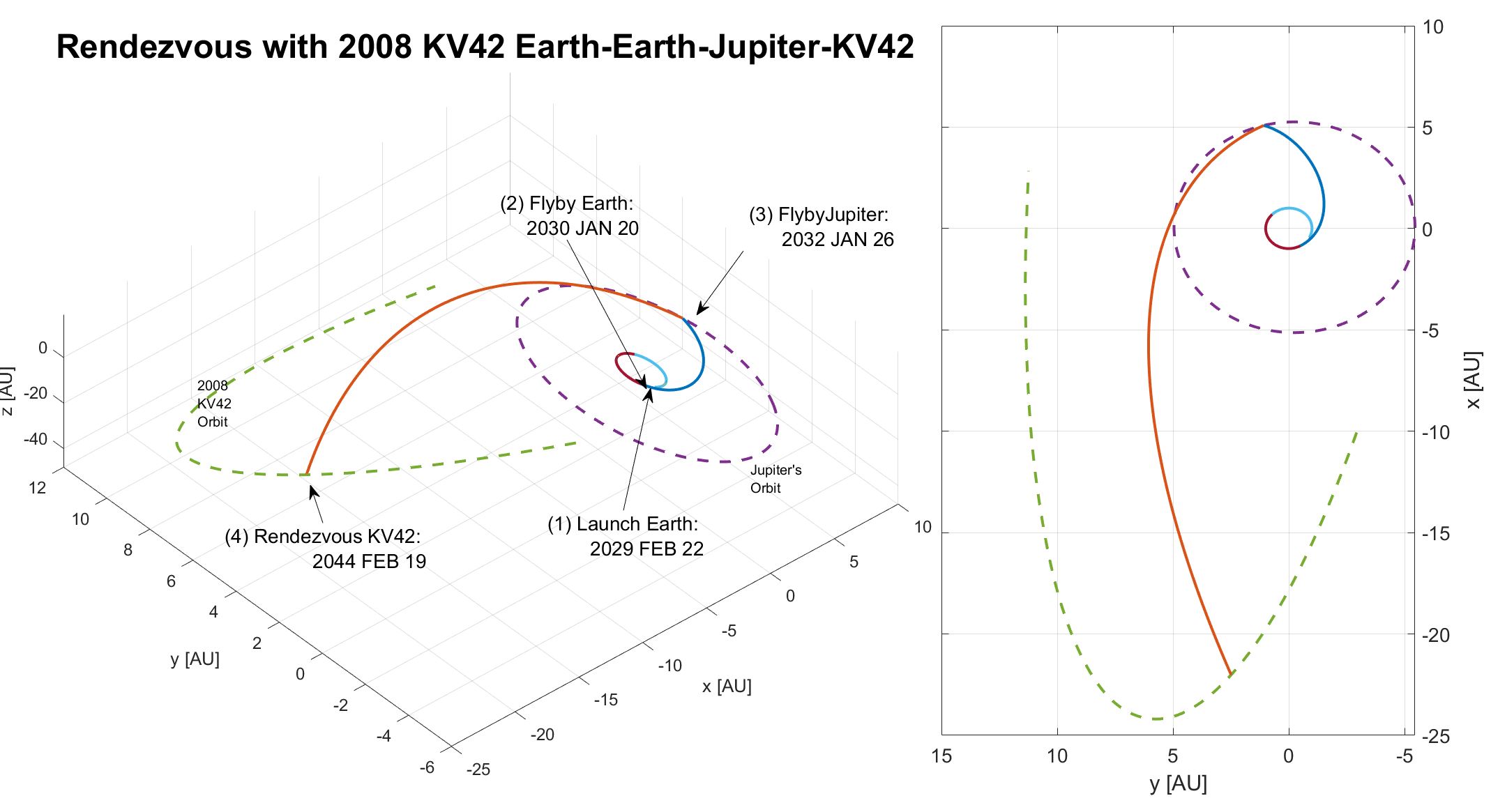}
\caption{Rendezvous Mission to KV42}
\label{fig:KV42}
\end{figure*}

\begin{figure*}
\centering
\includegraphics[scale=0.5]{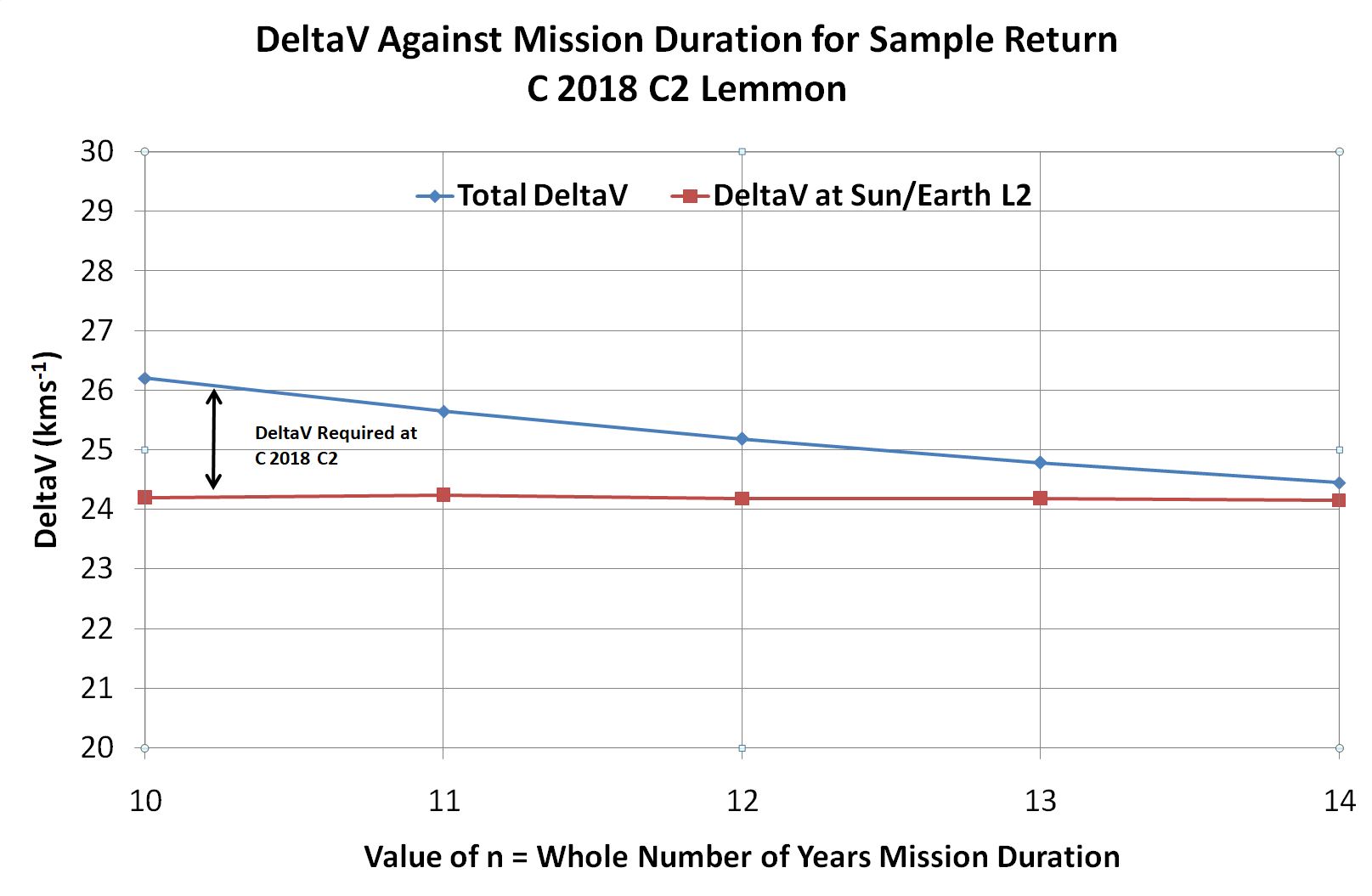}
\caption{Sample Return from C 2018/C2. $\delV$ Dependency on In-flight Time n (= Number of Years) }
\label{fig:2018C2_samp_ret_1}
\end{figure*}

To give an idea of the long term feasibility of performing a rendezvous missions with 514107, trajectories are provided for the years 2024 to 2038 in Figure \ref{fig:514107_3}. The upper blue line shows $\delV$s for missions without a 1 year $\infV$ Leveraging Maneuver, and revealing a periodicity of around 4 years between consecutive minima or maxima. If we take maxima or minima $\delV$ missions and introduce a preceding $\infV$ Leveraging Maneuver, we obtain the $\delV$ requirements indicated by the red squares below the blue line. Thus, a preceding $\infV$ Leveraging Maneuver can yield a reduction in $\delV$ of around 40\%. In the case of the highly inclined centaur 2008 KV42, a rendezvous mission seems feasible with a launch in 2029 and flight duration of 15 years from launch to rendezvous, see Figure \ref{fig:KV42} and Table \ref{table:KV42Rendez}.

\begin{table*}
\centering
\caption{Rendezvous Mission to 514107 (Possible Type 6 ISO)}
\label{table:514107Rendez}
\vspace{0.1 in}
\begin{tabular}{|c|c|c|c|c|c|c|c|}
\hline
\textbf{Number} & \textbf{Body} & \textbf{Time} & \textbf{Arrival} & \textbf{Departure} & \textbf{$\delV$} & \textbf{Cumulative} & \textbf{Periapsis}\\
{} & {} & {} & \textbf{speed} & \textbf{speed} & {} & \textbf{$\delV$} &  \textbf{altitude} \tabularnewline
\textbf{} &  &  & \si{km.s^{-1}} & \si{km.s^{-1}} & \si{km.s^{-1}} & \si{km.s^{-1}} & \si{km}\tabularnewline
\hline
\hline
\textbf{1} & Earth & 2030 MAR 04 & 0.00 & 0.01 & 0.01 & 0.01 & N/A\tabularnewline
\hline
\textbf{2} & Earth & 2031 MAR 02 & 0.01 & 15.88 & 8.31 & 8.32 & 200\tabularnewline
\hline
\textbf{3} & Jupiter & 2032 JAN 13 & 24.80 & 24.93 & 0.07 & 8.39 & 77197.6\tabularnewline
\hline
\textbf{4} & 514107 & 2039 NOV 06 & 2.49 & 0.00 & 2.49 & \textbf{10.88} & N/A\tabularnewline
\hline
\end{tabular}
\end{table*}

\begin{table*}
\centering
\caption{Rendezvous Mission to 2008 K4V2 (Possible Type 6 ISO)}
\label{table:KV42Rendez}
\vspace{0.1 in}
\begin{tabular}{|c|c|c|c|c|c|c|c|}
\hline
\textbf{Number} & \textbf{Body} & \textbf{Time} & \textbf{Arrival} & \textbf{Departure} & \textbf{$\delV$} & \textbf{Cumulative} & \textbf{Periapsis}\\
{} & {} & {} & \textbf{speed} & \textbf{speed} & {} & \textbf{$\delV$} & \textbf{altitude} \tabularnewline
\textbf{} &  &  & \si{km.s^{-1}} & \si{km.s^{-1}} & \si{km.s^{-1}} & \si{km.s^{-1}} & \si{km}\tabularnewline 
\hline
\hline
\textbf{1} & Earth & 2029 FEB 22 & 0.00 & 0.11 & 0.11 & 0.11 & N/A\\
{} & Launch & & & & & &\tabularnewline
\hline
\textbf{2} & Deep Space & 2029 AUG 07 & 29.68 & 29.65 & 0.59 & 0.69 & N/A\\
{} & Maneuver & & & & & &\\
{} & at 1.0AU & & & & & &\tabularnewline 
\hline
\textbf{3} & Earth & 2030 JAN 20 & 0.48 & 8.93 & 3.16 & 3.85 & 200\\
{} & Powered Flyby & & & & & & \tabularnewline
\hline
\textbf{4} & Jupiter Flyby & 2032 JAN 26 & 7.48 & 9.95 & 0.78 & 4.63 & 302223.2\tabularnewline 
\hline
\textbf{5} & 2008 KV42 & 2044 FEB 19 & 9.01 & 0.00 & 9.01 & \textbf{13.65} & N/A\tabularnewline 
\hline
\end{tabular}
\end{table*}

\subsection{Sample Return Missions}

With NTP, sample returns are feasible from type (1), (2) \& (4) ISOs, beginning with a pre-positioned interceptor loitering at the Sun/Earth L2 (SEL2) point, where the probe awaits a dispatch order upon detection of an ISO. Not all, but some, weakly hyperbolic comets have orbits appropriate for a direct return to Earth. A sample loiter/interceptor mission to C/2020 N1, serving as a surrogate for a type (2) object \& and possibly a type (4) ISO, is shown in Figure \ref{fig:C2020N1}. A future discovery of such an object would have an identical general sample return mission architecture to that shown but different values for mission duration, $\delV$ and launch date. 

When an ISO conducive to sample return is discovered, a heliocentric ellipse from Earth is computed. Requirements for this ellipse are (a) it intercepts the comet with relative velocity $<$ 6 \si{km.s^{-1}} (b) its time period is a whole number of $n$ years, (c) it minimizes $\delV$ required at SEL2 departure. Note that (b) ensures free return to Earth without any plane changes or any other $\delV$s along this ellipse. For the chosen target, the departure $\delV$ is applied at the optimal launch time using NTP or solar electric propulsion with arcjets. As the target is approached, an impactor is deployed and the spacecraft travels through the plume. If the plumes are anticipated to be hazardous (e.g., based on prior spectroscopy), a swarm of subprobes can be released and sent in advance of the main craft to sample the plume, returning to the main craft at a safe standoff distance after the encounter. 

The spacecraft arrives back at Earth for aerocapture and eventual return to Earth’s laboratories. For three currently known, weakly hyperbolic comets, which would have been suitable for this sort of sample return during their passage through the inner solar system (optimal launch dates have lapsed), $\delV$s is predicted to range from 17.4 \si{km.s^{-1}} to 24.4 \si{km.s^{-1}}, $n$ from 10-17 years, and intercept distance from 4.5-10\si{AU}. Using NTP, payload masses on the order of several metric tonnes are achievable, assuming the availability of an SLS Block 2 and two zero-boil-off and zero-leakage LH2 tanks, of the kind assumed in NASA’s Manned Mars Mission Design Reference Architecture,\fnref{fna} with optimal mass ratio. 

Table \ref{tab:samp_ret} gives a list of some more weakly hyperbolic comets used as surrogates for type (4) objects, but note that any of these also could be a type (2) ISO. It can be seen that three such objects are candidates for sample returns of the kind described, C 2020 N1 P, C 2018 C2 ‘Lemmon‘ and C 2014 Y1. If we constrain the spacecraft departure date to be after the discovery date (which was 28/01/2018), focus our intention, for example, on C 2018 C2, and use Optimum Interplanetary Trajectory Software to solve such trajectories,\fnref{fnb} there turns out to be several sample return solutions with different values of $n$, i.e. n = 10, 11, 12, 13 and 14 years. We also find that the departure date is always very close to 08/05/2018 and the thrust directions lie within around $\ang{1}$ of each other. The $\delV$ at departure stays just about constant as the value of $n$ increases, but there arises a gradually increasing $\delV$ at intercept. The combined effect is to increase the total $\delV$ requirement as $n$ increases. All this information is provided in the Figure \ref{fig:2018C2_samp_ret_1}. Note that this assumes a departure from the SEL2 point directly into the heliocentric ellipse, although a gravitational assist on Earth would possibly be more efficient.
\fntext[fna]{\url{https://www.nasa.gov/pdf/373665main_NASA-SP-2009-566.pdf}}
\fntext[fnb]{\url{https://github.com/AdamHibberd/Optimum_Interplanetary_Trajectory/blob/master/doc/Optimum\%20Interplanetary\%20Trajectory\%20Software\%20by\%20Adam\%20Hibberd.pdf}}

\subsection{Discussion of Mission Findings}
Missions to ISOs might resolve many vital questions about our and other star systems, are technologically feasible, but some mission types face noteworthy challenges regarding technology maturity. To be specific, it is expected that further development and deployment of heavy launcher and NTP systems would benefit the exploration of potential ISOs greatly.

Our results indicate that most mission types elucidated herein, except for sample return, could be realized with existing technologies or modified versions of existing technologies, such as chemical propulsion and a Parker Solar Probe-type heat shield \citep{Hein-et-al-2017-a,HHE20}. Collisions with dust, gas, and cosmic rays and spacecraft charging in the interplanetary or interstellar medium will engender deflection of the spacecraft trajectory and cause material damage to it, but both effects are likely minimal even at high speeds \citep{HLBL,HL17,LL20,LL21}, and the former can be corrected by onboard thrusters. However, for sample return missions, technologies which currently have a low Technology Readiness Level (TRL) would be required, such as NTP, for which TRL ranges from 2 to 6, depending on the reference (e.g., NASA Technology Taxonomy,\fnref[fnc] NASA Technology Roadmap), as well as zero-boil-off and zero-leakage LH2 tanks. 
\fntext[fnc]{\url{https://www.nasa.gov/offices/oct/taxonomy/index.html}}
Moreover, missions involving a Solar Oberth maneuver are particularly sensitive to uncertainties in the perihelion burn and might be difficult to accurately steer towards the ISO in actuality. Perihelion burn uncertainties are relevant for solid-propellant rockets. This issue may be particularly applicable to ISOs which are on their way out of the solar system, given the variability accompanying the position determination of such ISOs \citep{Hein-et-al-2017-a}. Hence, although the Solar Oberth maneuver accords considerable advantages in terms of performance, it still needs to be proven in practice.

As a consequence, for now, we are left with the conundrum of either waiting for the next ISO to be discovered via a loiter mission, to chase an ISO already on its way out of the solar system, or to develop NTP for facilitating access to a greater variety of ISOs. 

\section{Conclusions}
There are many mysteries that remain unresolved about the Solar system, which can be distilled down to a single question: Is the Solar system typical? In other words, does it obey the Copernican Principle sensu lato? The detection of exoplanets has, thus far, enabled us to address this issue to an extent insofar as the architecture and general makeup of planetary systems is concerned. However, we still remain in the dark when it comes to more specific questions such as the modality of planet formation, the composition and interior structure of rocky and/or icy objects, the gravitational ejection of planetesimals, and obviously the prevalence of prebiotic chemistry and life. It is apparent that a first-hand study of ISOs, along the lines proposed herein, may enable us to settle most, if not all, of these vital questions, thereby paving the way toward a more in-depth assessment of the Copernican Principle.

Hence, the goal of this paper was to explore whether missions to various categories of ISOs are realizable by utilizing existing or near-term technology. The answer is in the affirmative as illustrated by our analysis in Sec. \ref{SecISO}. Such near-term missions would generate in-situ data from bona fide extrasolar objects, the scientific value of which is difficult to overstate, without actually flying to other stellar systems. We presented concrete scenarios for the actualization of the fast flyby, rendezvous, and sample return mission categories. 

A combination of Falcon Heavy or SLS launch vehicles, chemical propulsion, and Parker Solar Probe-derived heatshield technology would be sufficient for fast flybys. When it comes to a rendezvous, solar electric propulsion ought also be incorporated to achieve the appropriate mission constraints. Lastly, in the case of sample return, NTP would be rendered necessary as well. In the event of sufficiently quick detection and launch of the spacecraft, we showed that all three categories could be implemented with reasonable flight durations of $\sim$10 years. The minimal suite of onboard instruments for answering the questions posed a couple of paragraphs earlier, about the origin of these objects, is a camera and mass spectrometer; we will not delve into it further as it falls outside the scope of this paper.


\end{document}